\documentclass[journal]{IEEEtran}
\usepackage{cite}
\usepackage{amsmath,amssymb,amsfonts}
\usepackage{algorithmic}
\usepackage{graphicx}
\usepackage{textcomp}
\usepackage{xcolor}
\usepackage[hyphens]{url}
\usepackage{hyperref}

\usepackage{colortbl}
\usepackage{booktabs}
\usepackage{tabularx}
\usepackage{listings}

\newcommand{\Description}[1]{}

\title{BenchmarkAnything: Agent-Driven Construction of Simulator-Ready Microarchitecture Benchmarks}

\author{
  Xiangfeng Sun\textsuperscript{1,$\dagger$},
  Ceyu Xu\textsuperscript{1,$\dagger$,*},
  Chen Bai\textsuperscript{2},
  and Yuan Xie\textsuperscript{1} \\[1ex]
  \textsuperscript{1}The Hong Kong University of Science and Technology,
  Hong Kong SAR, China \\
  \textsuperscript{2}Fudan University, Shanghai, China \\[0.5ex]
  \textit{xsunbv@connect.ust.hk, eeentropy@ust.hk, cbai@fudan.edu.cn,
  yuanxie@ust.hk}%
  \thanks{\textsuperscript{$\dagger$}Xiangfeng Sun and Ceyu Xu contributed equally to this work.}
  \thanks{\textsuperscript{*}Corresponding author.}
}

\begin{document}
\maketitle

\begin{abstract}

The selection of benchmark workloads is of paramount importance in computer architecture, as it establishes the yardstick against which architectural innovations are measured and guided.
Yet for decades, the SPEC benchmark suites, comprising merely tens of workloads, have been the de facto standard in academic architectural research, where they are frequently treated as a principal evaluation and optimization target.
When a suite this small is relied upon so heavily, it risks architectural overfitting; as our research and prior studies demonstrate, an overly narrow focus can mislead design decisions by overvaluing certain innovations, producing cores that excel on SPEC benchmarks yet underperform on broader, realistic workloads.

To mitigate this overfitting, adopting a large, comprehensive benchmark suite is the natural solution. However, the immense engineering effort required to strip software into the clean, interference-free binary executables demanded by simulators often makes this highly impractical.
In this work, we demonstrate that AI agents provide an elegant solution to this challenge. Rather than manually curating yet another static benchmark suite, we introduce an agent-driven workflow capable of autonomously transforming arbitrary open-source repositories into simulator-ready executables. 
This automated approach makes workload collection highly scalable, allowing us to rapidly harvest hundreds of diverse applications from public repositories into our benchmark suite. 
Through a comparative analysis of our agent-generated suite against SPEC, we show that it not only achieves higher-fidelity performance assessments but also uncovers novel architectural insights that traditional, static suites fail to expose.

\end{abstract}

\begin{IEEEkeywords}
Benchmarking, microarchitecture, workload characterization, LLM agents, computer architecture simulation.
\end{IEEEkeywords}

\section{Introduction}\label{sec:intro}
\begin{figure}[t]
    \centering
    \includegraphics[width=1\linewidth]{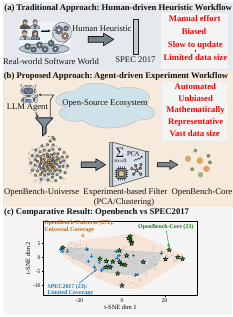}
    \caption{Comparison of human-driven heuristic workflow and agent-driven experiment workflow. (a) Traditional human heuristic curation is labor intensive and yields limited workload diversity. (b) The agent-driven workflow autonomously gathers applications to form the vast \textit{openbench-universe}. This pool is then statistically distilled into a highly efficient and representative subset named \textit{openbench-core}. (c) Projecting these workloads into the microarchitectural feature space reveals the limited coverage of legacy suites. In contrast, \textit{openbench} spans a much broader behavioral spectrum and better covers contemporary execution bottlenecks.}
    \Description{Three-panel figure comparing human-driven benchmark curation with an agent-driven workflow, and showing that openbench spans a broader microarchitectural feature space than legacy suites.}
    \label{fig:intro}
\end{figure}

Much of computer architecture research begins with bottlenecks revealed by benchmark workloads and ends with evaluation on benchmark suites. In practice, however, architects cannot make design decisions with full visibility into the breadth of real-world deployment scenarios. Human attention is limited, simulation is expensive, and many important production workloads are inaccessible to researchers. As a result, the community relies on curated benchmark suites that aim to represent broad classes of real-world applications in a principled and reproducible manner. Suites such as SPEC therefore serve as essential proxies for hardware research and architectural innovation.

Because benchmark construction is so consequential, the community has invested decades of expert effort into maintaining recognized benchmark suites that attempt to identify the workloads most representative of real-world use. Yet despite this sustained effort, the combined SPEC family still contains fewer than one hundred benchmark cases across all of its major versions. This limited scale raises a fundamental question for architecture research: can design decisions derived from such a small benchmark set be sufficiently reliable and representative of pervasive real-world workloads?

Prior work has repeatedly shown that SPEC is not fully representative of contemporary real-world workloads and can therefore lead to suboptimal architectural conclusions. Even when SPEC remains informative for certain workload classes, the broader benchmark construction paradigm is still in crisis. The root cause is that benchmark curation has long relied on a human-driven, heuristic-based process. Because the real-world workload space is effectively unbounded while expert attention is inherently limited, benchmark designers must select a small initial candidate pool based largely on subjective judgments of importance and representativeness. This manual selection introduces inherent bias and creates severe representativeness limitations. It is also expensive, labor intensive, and fundamentally difficult to scale. In principle, benchmark construction would benefit from starting from as large and diverse a workload pool as possible and then using a data-centric methodology to distill representative subsets. In practice, however, manual engineering effort sharply limits the size, freshness, and diversity of the candidate pool that can be considered. Moreover, different researchers, companies, and institutions often care about different workload mixes and deployment scenarios. A single static benchmark suite cannot adapt to these varying needs, nor can it evolve fast enough to track the rapid emergence of new applications and software stacks. Taken together, these limitations indicate that the field needs a paradigm shift in benchmarking toward methodologies that are more representative, less biased, more scalable, and more transferable across use cases.

To overcome the fundamental limitations of the human-driven, heuristic-based benchmark construction paradigm, we propose a shift toward an agent-driven, data-centric benchmark construction methodology that is more representative, less biased, more scalable, and more transferable across use cases, as shown in Figure~\mbox{\ref{fig:intro}}. Specifically, we introduce \textit{BenchmarkAnything}, a fully automated framework designed to improve workload representativeness through large-scale workload collection and statistical distillation. This methodology operates in two phases. In the first phase, an LLM-driven agent flow autonomously navigates open-source platforms to collect candidate applications, resolve build dependencies, and compile them into executable artifacts. This automated collection process reduces human selection bias and scales with the pace of software evolution. In the second phase, the framework executes the collected applications to extract a broad set of microarchitectural features. It then applies rigorous experimental and statistical analysis to systematically distill this large candidate pool into a benchmark suite with strong empirical representativeness over the collected workload space.

Based on this agent-driven, data-centric benchmark construction methodology, we construct the \textit{openbench} benchmark family. It includes \textit{openbench-universe}, a large benchmark suite that bridges industry-standard tasks with cutting-edge software deployments, and \textit{openbench-core}, a distilled subset for efficient simulation. \textit{openbench-universe} consists of 272 benchmarks spanning 14 domains, providing substantial diversity and coverage. Our comprehensive microarchitectural characterization (\mbox{\ref{Characterization}}) shows that \textit{openbench} achieves higher fidelity than legacy alternatives. By reducing manual selection bias and grounding benchmark selection in experimental validation, \textit{openbench} more faithfully captures the memory-hierarchy pressure and instruction footprints of contemporary workloads \cite{Hassan2021A, Navarro-Torres2019Memory, Wang2014BigDataBench}. As a result, it exposes critical architectural bottlenecks that legacy benchmark suites often fail to trigger. Overall, \textit{openbench} provides a more realistic and timely evaluation target for the design of next-generation high-performance processors. Our aim is to \emph{complement} rather than replace SPEC: SPEC remains a carefully curated, stable, high-quality reference, while \textit{BenchmarkAnything} contributes a fast, low-cost, and portable way to keep evaluation aligned with software as it evolves. Because the framework is retargetable, the same flow can convert a research group's or a company's own important applications into a simulator-ready suite, rather than committing every user to a single fixed set of workloads.

In summary, this paper makes the following contributions.
\begingroup
\setlength{\leftmargini}{1.2em}
\begin{itemize}
    \item We introduce an agent-driven, data-centric benchmark construction paradigm that replaces traditional manual heuristics with large-scale automated workload collection and statistical distillation.
    \item We present \textit{BenchmarkAnything}, a fully automated framework that can autonomously search open-source platforms, resolve build dependencies, and produce simulator-ready benchmark artifacts.
    \item We construct the \textit{openbench} benchmark family, including its universe and core subsets, and show that it exposes modern execution bottlenecks that are largely absent in SPEC CPU2017.
    \item We show that this workflow substantially broadens workload diversity and coverage while preserving simulation efficiency through principled core-subset selection.
\end{itemize}
\endgroup

%
%

\section{Background and Motivation}\label{sec:bg}
\subsection{Benchmark Importance for CPU Design}
Benchmarks are a primary feedback proxy between real software and CPU design decisions. Because processor performance depends jointly on microarchitecture and workload characteristics, architects rely on benchmark suites to approximate the target deployment space during pre-silicon design, post-silicon evaluation, and subsequent tuning of compilers and runtimes \cite{Wang2019Predicting,6704671}. A well-designed suite provides a common, repeatable basis to compare alternative designs, quantify bottlenecks, and reason about tradeoffs in performance, power, and area \cite{Wang2024A,Buduleci2024Architectural}. The choice and construction of benchmarks therefore strongly affect the validity of design conclusions. Navigating this space, however, is fundamentally bottlenecked by simulation overhead: simulating very large application sets is infeasible, so architects must select representative benchmark and input pairs.

\subsection{Existing CPU Benchmark and Manual Benchmark Curation}

Standardized suites such as SPEC CPU have been central to microarchitecture research and industrial CPU evaluation for more than three decades \cite{Hassan2021A,Kounev2020The,Singh2019Memory}. SPEC suites stress the processor core, memory hierarchy, and compiler toolchain under a controlled methodology, and have become the de facto basis for comparing commercial systems and validating academic proposals. The latest release, SPEC CPU2017, extends CPU2006 with larger and more complex integer, floating-point, and multi-threaded workloads to better reflect contemporary software \cite{Limaye2018A,Panda2018Wait,Hassan2021A}. All of these suites rely on significant manual curation: selecting applications, defining build procedures, resolving dependencies, and engineering representative inputs under strict runtime and resource constraints.

Despite their impact, existing CPU benchmarks and manual curation exhibit several limitations. First, SPEC CPU2017 shows substantial internal self-similarity and does not fully capture emerging workload classes. Performance counter and clustering studies report strong redundancy and show that most CPU2017 workloads occupy a narrow region of the behavior space, while many modern data center, graph analytics, and EDA workloads exhibit distinct characteristics \cite{Wang2019Predicting,Song2018Experiments,Panda2018Wait}. Memory hierarchy studies find that almost half of SPEC CPU2006 and CPU2017 benchmarks have very low last-level cache miss ratios on contemporary processors, and that CPU2017 places even less pressure on the memory hierarchy than CPU2006 \cite{Hassan2021A, Navarro-Torres2019Memory}. Traditional CPU suites also show lower L1 instruction miss rates and higher instruction locality than big data and internet service workloads, limiting their ability to stress front-end structures and large caches realistically \cite{Wang2014BigDataBench}. SPEC-based evaluations may therefore systematically underrepresent the instruction footprints, working set sizes, and microarchitectural bottlenecks that dominate modern cloud and data-intensive software.

Second, manual curation does not scale with the size and dynamism of modern software ecosystems. The creation of SPEC CPU2017 required more than a decade, and the suite still ships with only a small number of official input sets per benchmark \cite{Amaral2018The}. Amaral et al. note that the lack of multiple workloads per application complicates evaluation of techniques that depend on workload variation, and that constructing additional workloads demands substantial manual effort and deep application knowledge. Similar concerns arise in achieving consistent and comparable CPU2017 evaluation across configurations, where arbitrary specificity in some parameters and ambiguity in others lead to large performance variability and unfair comparisons \cite{Wang2024Achieving}. More broadly, benchmarking and benchmark maintenance are increasingly recognized as labor-intensive, error-prone processes that require careful workflow management and automation to remain reproducible and extensible \cite{Toprak2025Employing,Mallona2024Building}. Crucially, the persistent lack of a standardized, automated methodology prevents the community from scaling benchmark suites to match the rapid evolution of modern software.

\subsection{LLM Agents in Software Engineering}
Recent advances in large language models have catalyzed a shift from passive code models to LLM-based agents that perceive context, invoke tools, and autonomously plan multi-step workflows for software engineering tasks \cite{He2024LLM-Based, Liu2024Large, Jin2024From,Wang2024Agents,Hou2023Large,Cheng2024Exploring}. Compared with single shot LLM usage, these agents aim to cover broader portions of the software development lifecycle and to behave more like collaborative team members than isolated prediction engines \cite{Applis2025Unified, Zhang2024AutoCodeRover}.

Surveys now distinguish three layers of capability. First, LLMs as code models support code completion, translation, summarization, and repair \cite{Zheng2023Towards,Wang2023Software,Ross2023The,Weber2024Significant}. Second, LLM-based agents wrap LLMs with perception, memory, and action modules to enable tool use, repository-level reasoning, and iterative refinement over long contexts. Third, multi-agent systems coordinate specialized agents that play roles such as designer, coder, tester, and reviewer \cite{Applis2025Unified}. Systematic reviews report rapid growth of this line of work across requirements engineering, code generation, test generation, maintenance, and DevOps \cite{Jin2024From,Radenkovic2025Application}. Recent open-source systems such as CLI-Anything make this shift concrete by transforming arbitrary repositories into agent-native CLIs with structured outputs and stable command interfaces \cite{CLIAnything2026}. Our aim of abstraction is different: rather than turning anything into a CLI, we turn diverse real-world software into benchmarks, a promising direction for addressing this long-standing scalability crisis.

\section{BenchmarkAnything: Agent-Driven Framework}\label{sec_openbench_framework}
\begin{figure*}[t]
    \centering
    \includegraphics[width=\textwidth]{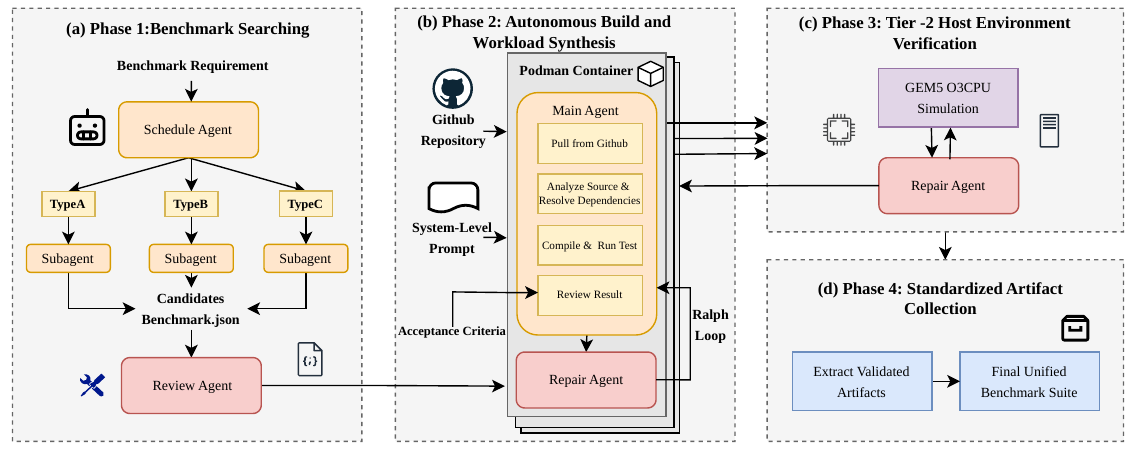}
    \caption{Overview of the \textit{BenchmarkAnything} framework. (a) Phase~1, Benchmark Searching. A scheduling agent classifies benchmark requirements into Type~A, Type~B, and Type~C, and dispatches dedicated subagents that retrieve candidate repositories from open-source platforms in parallel. (b) Phase~2, Autonomous Build and Workload Synthesis. Within an isolated Podman sandbox, the main agent resolves software dependencies and performs architecture-aware compilation. Compilation or execution failures trigger the Ralph loop, an iterative reflection-and-repair cycle where a repair agent diagnoses errors and supplies targeted feedback until the benchmark satisfies strict static-linking, execution, and output acceptance criteria. (c) Phase~3, Tier-2 Host-Environment Verification. Compiled artifacts undergo microarchitectural compatibility validation through gem5 simulation. Simulation failures first invoke a repair agent for targeted fixes; unresolvable issues are routed back to the container for recompilation. (d) Phase~4, Standardized Artifact Collection. The framework extracts validated binaries, wrapper scripts, and input data, and reorganizes them into a uniformly formatted, simulation-ready benchmark suite.}
    \Description{Overview diagram of the BenchmarkAnything workflow, showing benchmark searching, autonomous build and workload synthesis, host-environment verification, and standardized artifact collection.}
    \label{fig:agent_flow}
\end{figure*}

Our key insight is that \emph{any} open-source software repository can be automatically transformed into a standard benchmark through a sufficiently capable autonomous agent workflow. The open-source ecosystem continuously generates software that reflects contemporary computing patterns, yet manual curation cannot harvest this diversity at scale. To bridge this gap, we propose \textit{BenchmarkAnything}, an automated framework that leverages LLM-based agents to autonomously retrieve, compile, verify, and standardize benchmark applications from diverse open-source repositories. Figure~\ref{fig:agent_flow} presents the end-to-end workflow, which comprises four phases: benchmark searching, autonomous build and workload synthesis within an isolated sandbox, host-environment verification through microarchitectural simulation, and standardized artifact collection. In our current implementation, 93\% of the open-source projects retrieved in Phase~1 are successfully transformed into standardized benchmarks by this end-to-end flow. All agents are driven by a single backbone model, MiniMax-M2.5, which balances cost efficiency, throughput, and coding proficiency. The remainder of this section describes each phase and its design rationale.

\subsection{Phase 1: Benchmark Searching}
The open-source ecosystem hosts millions of repositories spanning diverse languages, build systems, and application domains, so identifying candidates requires a search strategy that is both broad and systematic. As illustrated in Figure~\ref{fig:agent_flow}(a), the first phase analyzes the user-provided benchmark requirements.
A primary scheduling agent classifies them into distinct categories and dispatches each category to a dedicated subagent.
The categories encode \emph{how} a candidate should be sought. \textbf{Type~A} requirements are domain- or algorithm-driven, such as graph analytics or sparse linear algebra. \textbf{Type~B} requirements are adoption-driven and target widely used, actively maintained projects such as high-star repositories or packages with large install bases. \textbf{Type~C} requirements are capability- or keyword-driven, such as a compression codec or a regular-expression engine. A single user request typically expands into a mixture of all three categories, so the candidate pool spans both flagship applications and the long tail of the ecosystem.
This parallelized decomposition is deliberate: different domains reside in distinct repository ecosystems with varying naming conventions and metadata structures, and specialized subagents apply domain-appropriate search heuristics more effectively than a single monolithic query.

Search and review are governed by explicit, hardware-aware selection criteria. Each candidate is scored along four axes. \emph{Adoption and relevance} captures repository popularity, release maturity, active maintenance, and documentation quality. \emph{CPU-centric executability} requires a real application entry point that performs substantive computation on the CPU. \emph{Input availability} prefers official or repository-provided inputs and documented public datasets. \emph{Simulator compatibility} requires the workload to be admissible under the target simulator. The review agent additionally enforces an exclusion filter: repositories that are GPU-only, accelerator-only, or expressed purely in CUDA/OpenCL/HLS, as well as projects containing only unit tests, micro-demos, or build scaffolding, are rejected unless an explicit CPU implementation is targeted.
The retrieved results are aggregated into a unified \texttt{CandidatesBenchmark.json} manifest, which a review agent evaluates to eliminate duplicate and non-compliant entries.
If the number of valid candidates falls below the target threshold, the review agent instructs the search agent to retrieve more. In practice the search agent deliberately over-provisions candidates, because collecting a large initial pool and filtering offline is far more efficient than issuing repeated narrowly scoped queries, so secondary searches are rarely needed.
Once validated, the manifest is forwarded to Phase~2.

\begin{figure}[htbp]
\centering
\lstset{
    basicstyle=\ttfamily\scriptsize,
    backgroundcolor=\color{gray!10},
    frame=single,
    framerule=0.5pt,
    rulecolor=\color{gray!80},
    breaklines=true,
    captionpos=b,
    showstringspaces=false,
    numbers=none
}
\begin{lstlisting}[caption={Extracted rules and constraints from the \textit{main agent} prompt.}, label={lst:prompt_listing}]
[Role] Autonomous executor for BenchmarkAnything with sudo access.

[Step 1] Fetch source code from GitHub.
[Step 2] Compile with STRICT rules:
  - Install missing deps autonomously.
  - MUST enforce static linking.
  - Use `-march=x86-64`. PROHIBIT AVX/AVX2/AVX-512 to ensure gem5 SE compatibility.
  ...
[Step 3] Prepare Inputs & Run Script:
  - Use official or repository-provided inputs.
  - Generate `run.sh`. No infinite loops or sleep commands.
  ...
[Acceptance Criteria]
  1. Binary is a statically linked ELF.
  2. Native execution succeeds (exit code 0).
  3. Retires enough instructions to fill the 1B-insn warmup + 100M-insn ROI window.
\end{lstlisting}
\Description{Listing of the main-agent prompt constraints, including repository fetching, strict static compilation, and benchmark acceptance criteria.}
\end{figure}
\subsection{Phase 2: Autonomous Build and Workload Synthesis}

Compiling arbitrary open-source software is inherently unpredictable. Each repository may depend on a unique combination of libraries, toolchains, and system utilities, and resolving these dependencies on a shared host risks polluting the build environment of subsequent candidates. Figure~\ref{fig:agent_flow}(b) illustrates how the second phase addresses this challenge.
The framework employs Podman to instantiate an isolated container for each candidate, eliminating the need for superuser privileges on the host while enabling safe resolution of heterogeneous dependencies.
Within the container, the agent is granted elevated privileges strictly for invoking package managers and installing missing libraries.
A system-level prompt governs compilation, enforcing output requirements such as a statically linked x86-64 executable, because statically linked binaries are a strict prerequisite for simulated execution environments that lack dynamic linker support.

The main agent drives the core build process.
It pulls the source code, analyzes the project documentation to resolve dependencies, and performs architecture-aware compilation that disables unsupported vector extensions such as AVX-512 and enforces single-threaded execution.
The agent then provisions input datasets following a strict provenance hierarchy: it first prefers official or repository-provided inputs and the project's own example workloads, falling back to documented public datasets referenced by the project. The chosen source and its provenance are recorded in the benchmark manifest, and when a binary exposes distinct behavior under different inputs the agent may retain multiple input configurations. Each input is sized so that the workload retires enough instructions to fill the subsequent 1B-instruction warmup and the full 100M-instruction region of interest, guaranteeing a measurable steady-state window rather than a setup-dominated run.
After initial native execution tests, the framework evaluates the result against three formal acceptance criteria. If any criterion fails, the workload enters the Ralph loop, an iterative reflection-and-repair mechanism connecting the repair agent and the main agent.
In each iteration, the repair agent diagnoses compilation logs or runtime errors and provides targeted feedback, enabling the main agent to adjust dependencies, compilation flags, or input configurations. The cycle continues until all criteria hold or a configurable iteration budget is exhausted.
Three criteria gate the exit: the binary must be a strictly statically linked ELF executable, it must run through a standardized shell script without crash signals such as segmentation faults, and it must retire enough instructions natively to fill the 1B-instruction warmup and 100M-instruction region of interest, a window whose completion is then confirmed under gem5 in Phase~3.
Workloads that satisfy all three proceed to the next phase.

\subsection{Phase 3: Tier-2 Host Environment Verification}

Passing the intra-container criteria ensures the benchmark works as a native executable but does not guarantee compatibility with a detailed microarchitectural simulator. System calls, memory-mapped I/O, and complex vector instructions that work on native hardware may cause fatal errors in simulation, so a second verification tier is essential. As illustrated in Figure~\ref{fig:agent_flow}(c), the compiled artifacts cross the isolation boundary into the host environment, where the framework executes each workload under a gem5 O3CPU simulation.
If the simulation fails due to unsupported system calls or unrecognized instructions, a repair agent first applies targeted fixes directly to the execution artifacts. If the repair agent cannot resolve the issue, a fallback strategy routes the workload back to the main agent in the container for complete recompilation under stricter architectural constraints.

\subsection{Phase 4: Standardized Artifact Collection}

The final phase, shown in Figure~\ref{fig:agent_flow}(d), addresses a persistent barrier to benchmark adoption. When researchers manually port open-source workloads for simulation, they must contend with heterogeneous build artifacts, inconsistent input formats, and undocumented execution scripts, an overhead that often discourages adopting new workloads entirely.
Our framework instead extracts only the validated artifacts from the sandbox, discarding intermediate build files and isolating the core functional components: the statically linked binary, the execution wrapper script, and any required input data.
By enforcing this uniform directory topology across all applications, the framework yields a plug-and-play suite that deploys seamlessly in subsequent architectural experiments.
This phase also makes the suite reproducible despite LLM nondeterminism. \textit{BenchmarkAnything} separates \emph{discovery} from \emph{replay}: agent decisions may differ across runs, but each accepted benchmark freezes the source revision, build commands and flags, input files and provenance, execution command, validation checks, and gem5 invocation. The published package can therefore be replayed without invoking a model and remains fixed even if discovery yields another candidate universe.

\subsection{The Ralph Loop: Iterative Reflection and Repair}

Among the four phases, the Ralph loop is the single most critical mechanism that enables \textit{BenchmarkAnything} to operate at scale without human intervention. The Ralph loop is a general agentic pattern in which an agent is repeatedly re-prompted against a fixed specification until it succeeds, originally articulated by Geoffrey Huntley~\cite{huntley2025ralph} and since widely adopted in agentic coding frameworks. Our contribution is not the loop itself but its specialization to benchmark construction: a formal, simulator-aware acceptance gate comprising static-linking, crash-free execution, and region-of-interest completion checks that decides when an iteration may exit.

The challenge is fundamental. Unlike curated suites whose build procedures are hand-tuned over months, each collected repository presents a unique combination of build-system conventions, dependency graphs, and implicit environmental assumptions, so a single-pass compilation strategy would fail on a large fraction of candidates. The Ralph loop structures the build process as a closed-loop control system in which failures drive iterative refinement rather than immediate rejection.
In practice, static linking is the dominant engineering obstacle: many modern projects assume dynamic linking or pull in components such as NSS, network, or threading libraries that resist a fully static build. Substituting static-friendly libraries, stubbing optional features, and pruning irreducibly dynamic dependencies constitute the principal repair tasks the loop automates, and applications that cannot be reduced to a self-contained static binary are deferred by the acceptance gate.

Two properties deserve emphasis. First, the acceptance criteria in Listing~\ref{lst:prompt_listing} serve as a formal verification gate rather than a soft heuristic, ensuring that every benchmark entering subsequent phases is simulation-ready by construction. Second, the iterative repair cycle mirrors the debugging workflow of a human developer who reads error messages, hypothesizes a fix, and retries, but operates at machine speed and scales to hundreds of repositories in parallel.

\section{\textit{openbench} Characterization vs.\ SPEC CPU2017}\label{Characterization}
\subsection{Experimental Setup and Profiling Space}

To characterize microarchitectural behaviors rigorously, we evaluate 30 unique configurations detailed in Table~\ref{tab:configs}. This multidimensional design space encompasses commercial processors ranging from ultra-low-power embedded controllers to aggressive server-class out-of-order engines, and we prune infeasible design points to maintain architectural realism: pairing a wide 10-instruction fetch width with a simple in-order core, for example, is unbalanced and unrealistic. We scale execution pipelines by varying fetch widths from one to eight instructions per cycle across in-order and out-of-order models, and scale the reorder buffer from 64 to 384 entries with proportional issue queues to capture the diminishing returns of backend expansion. We stress application working sets by varying L1 caches from 8kB to 64kB and L2 caches from 64kB to 2MB, instantiate TAGE, Tournament, BiMode, and Local branch predictors to evaluate frontend control-flow sensitivity, and pair Signature Path and Stride prefetchers with DRRIP, SHiP, BRRIP, and LRU replacement policies to expose heterogeneous memory access patterns. We execute this profiling space with the gem5 simulator in Syscall Emulation mode under the x86 ISA. To avoid distortion from unrepresentative setup phases, we fast-forward the first 1B instructions of each benchmark as warmup on a simple CPU with warm caches, then reset all statistics and collect them over a precisely defined 100M-instruction Region of Interest (ROI) executed on the detailed out-of-order core. Both the 1B warmup and the 100M ROI are user-adjustable instruction counts of our flow and follow common practice in simulation studies~\cite{ainsworth2020muontrap}. Because any microarchitecture-dependent characterization can be biased by the chosen machine configurations~\cite{hoste2007microarchitecture}, our selection is deliberately spread across the principal axes, namely core type, pipeline width, reorder-buffer and load/store-queue depth, L1/L2 capacity, and the choice of branch predictor, prefetcher, and replacement policy, so the 30 points sample balanced, realizable designs rather than a dense grid of one structure in isolation. This many-parameter, machine-dependent sweep complements two established methodologies: statistically rigorous design-of-experiment sampling such as Plackett--Burman, which identifies a minimal set of configurations to probe a large space~\cite{yi2003statistically}, and broad multi-ISA and compiler similarity studies of benchmark suites~\cite{phansalkar2007analysis}.

\begin{table}[htbp]
\centering
\caption{Microarchitecture configurations used in the evaluation. We systematically explored various architectural configurations and pruned infeasible design points to cover a representative range of contemporary single-core CPU microarchitectures. Pipeline width refers to the unified fetch/decode/issue/commit width. For in-order cores, ROB and IQ are not applicable (-). L1 caches are symmetric in size and associativity (I/D). Prefetchers, Branch Predictors (BP), and Replacement (Repl) policies leverage built-in gem5 models.}
\label{tab:configs}
\small
\setlength{\tabcolsep}{3pt}
\resizebox{\columnwidth}{!}{%
\begin{tabular}{@{}llcccccllll@{}}
\toprule
\textbf{Cfg} & \textbf{Core} & \textbf{Width} & \textbf{ROB} & \textbf{IQ} & \textbf{LQ/SQ} & \textbf{L1 (I/D)} & \textbf{L2} & \textbf{BP} & \textbf{Prefetch} & \textbf{Repl} \\
\midrule
\textbf{C1}  & OoO & 8 & 384 & 128 & 96/72 & 64kB/8-way & 2MB/16-way  & TAGE   & SPP    & DRRIP \\
\textbf{C2}  & OoO & 8 & 384 & 128 & 96/72 & 64kB/8-way & 2MB/16-way  & TAGE   & Stride & SHiP \\
\textbf{C3}  & OoO & 8 & 256 & 97  & 72/56 & 32kB/8-way & 1MB/16-way  & TAGE   & Stride & BRRIP \\
\textbf{C4}  & OoO & 8 & 384 & 128 & 96/72 & 64kB/8-way & 256kB/8-way & TAGE   & None   & LRU \\
\textbf{C5}  & OoO & 8 & 384 & 128 & 96/72 & 64kB/8-way & 2MB/16-way  & Local  & Stride & DRRIP \\
\textbf{C6}  & OoO & 6 & 256 & 97  & 72/56 & 64kB/8-way & 2MB/16-way  & TAGE   & SPP    & SHiP \\
\textbf{C7}  & OoO & 6 & 192 & 80  & 56/42 & 32kB/8-way & 1MB/16-way  & Tourn. & Stride & DRRIP \\
\textbf{C8}  & OoO & 6 & 192 & 80  & 56/42 & 32kB/8-way & 1MB/16-way  & TAGE   & Stride & TreePLRU \\
\textbf{C9}  & OoO & 6 & 256 & 97  & 72/56 & 16kB/4-way & 2MB/16-way  & TAGE   & None   & BRRIP \\
\textbf{C10} & OoO & 6 & 128 & 64  & 48/36 & 64kB/8-way & 1MB/16-way  & Tourn. & Stride & LRU \\
\textbf{C11} & OoO & 4 & 192 & 80  & 64/48 & 64kB/8-way & 2MB/16-way  & TAGE   & SPP & SHiP \\
\textbf{C12} & OoO & 4 & 192 & 80  & 64/48 & 32kB/8-way & 1MB/16-way  & TAGE   & Stride & DRRIP \\
\textbf{C13} & OoO & 4 & 128 & 64  & 48/36 & 32kB/8-way & 512kB/8-way & Tourn. & Stride & TreePLRU \\
\textbf{C14} & OoO & 4 & 128 & 64  & 48/36 & 32kB/8-way & 512kB/8-way & BiMode & Stride & LRU \\
\textbf{C15} & OoO & 4 & 96  & 48  & 40/32 & 32kB/8-way & 256kB/8-way & Tourn. & None   & LRU \\
\textbf{C16} & OoO & 4 & 192 & 80  & 64/48 & 64kB/8-way & 1MB/16-way  & Local  & Stride & BRRIP \\
\textbf{C17} & OoO & 4 & 128 & 64  & 48/36 & 16kB/4-way & 512kB/8-way & TAGE   & Stride & Random \\
\textbf{C18} & OoO & 4 & 128 & 64  & 48/36 & 64kB/8-way & 2MB/16-way  & BiMode & None   & SHiP \\
\textbf{C19} & OoO & 3 & 96  & 48  & 40/32 & 32kB/8-way & 512kB/8-way & Tourn. & Stride & LRU \\
\textbf{C20} & OoO & 3 & 96  & 48  & 40/32 & 16kB/4-way & 256kB/8-way & BiMode & None   & TreePLRU \\
\textbf{C21} & OoO & 2 & 96  & 48  & 32/24 & 32kB/8-way & 512kB/8-way & BiMode & Stride & LRU \\
\textbf{C22} & OoO & 2 & 64  & 32  & 24/20 & 32kB/4-way & 256kB/8-way & BiMode & None   & LRU \\
\textbf{C23} & OoO & 2 & 64  & 32  & 24/20 & 16kB/4-way & 128kB/8-way & Local  & None   & FIFO \\
\textbf{C24} & OoO & 2 & 128 & 64  & 48/36 & 64kB/8-way & 1MB/16-way  & TAGE   & SPP    & DRRIP \\
\textbf{C25} & InO & 2 & -   & -   & -/-   & 64kB/8-way & 1MB/16-way  & Tourn. & Stride & LRU \\
\textbf{C26} & InO & 2 & -   & -   & -/-   & 32kB/4-way & 512kB/8-way & BiMode & Stride & LRU \\
\textbf{C27} & InO & 2 & -   & -   & -/-   & 16kB/4-way & 256kB/8-way & Local  & None   & LRU \\
\textbf{C28} & InO & 1 & -   & -   & -/-   & 32kB/4-way & 256kB/8-way & BiMode & None   & LRU \\
\textbf{C29} & InO & 1 & -   & -   & -/-   & 16kB/4-way & 128kB/8-way & Local  & None   & FIFO \\
\textbf{C30} & InO & 1 & -   & -   & -/-   & 8kB/2-way  & 64kB/4-way  & None   & None   & Random \\
\bottomrule
\end{tabular}%
}
\end{table}

\subsection{The Scale, Diversity, and Coverage of the \textit{openbench-universe}}

Equipped with these configurations, we evaluate the workload pool collected by \textit{BenchmarkAnything}, a superset we denote \textit{openbench-universe} containing 272 standardized applications from diverse open-source projects. Because \textit{BenchmarkAnything} can continuously collect more benchmarks, we focus on 272 for tractability. To make the suite transparent and auditable, Table~\ref{tab:workloads} lists workloads in each application domain. The 272 benchmarks will be open-sourced with the artifact when the paper is accepted.

\begin{table}[t]
\centering
\caption{Overview of the \textit{openbench-universe}: 272 automatically constructed benchmarks grouped into 14 application domains.}
\label{tab:workloads}
\small
\setlength{\tabcolsep}{4pt}
\renewcommand{\arraystretch}{1.18}
\begin{tabularx}{\columnwidth}{@{}>{\raggedright\arraybackslash}p{2.35cm} X@{}}
\toprule
\textbf{Domain} & \textbf{Benchmarks} \\
\midrule
\textbf{Numerical / HPC} (54) & OpenBLAS, FFTW3, blis, GSL, flint, sundials, zfp, \ldots \\
\textbf{Cryptography} (26) & mbedtls, botan, cryptopp, bearssl, Nettle, libtomcrypt, BLAKE2, monocypher, hacl-star, XKCP, \ldots \\
\textbf{Miscellaneous} (25) & hyperscan, QuantLib, SEAL, primesieve, TA-Lib, \ldots \\
\textbf{System / Embedded} (20) & sqlite, fio, mimalloc, jemalloc, tcmalloc, busybox, \ldots \\
\textbf{Parsing / String} (20) & pcre2, HarfBuzz, rapidjson, protobuf-c, pugixml, utf8proc, cmark, snowball, \ldots \\
\textbf{Data Structures} (18) & CRoaring, marisa-trie, judy, \ldots \\
\textbf{Graphics / Sim} (18) & bullet3, box2d-v3, pbrt-v4, appleseed, meshoptimizer, reactphys3d, \ldots \\
\textbf{Compression} (16) & brotli, fast-lzma2, lzfse, miniz, xxHash, streamvbyte, fpzip, \ldots \\
\textbf{Sorting} (16) & pdqsort, timsort, ska\_sort, \ldots \\
\textbf{Interp. / Compiler} (15) & lua, luau, mruby, duktape, jerryscript, chibi-scheme, asmjit, re2c, rv32emu, \ldots \\
\textbf{AI / Game} (11) & Stockfish, Fairy-SF, cadical, picosat, \ldots \\
\textbf{Graph / Search} (11) & igraph, snap, lemon, \ldots \\
\textbf{Bioinformatics} (11) & minimap2, htslib, kallisto, seqan, edlib, parasail, WFA2-lib, seqtk, \ldots \\
\textbf{Media / Signal} (11) & libwebp, libtiff, lame, giflib, speexdsp, \ldots \\
\bottomrule
\end{tabularx}
\end{table}

\begin{figure}[htbp]
    \centering
    \includegraphics[width=\columnwidth]{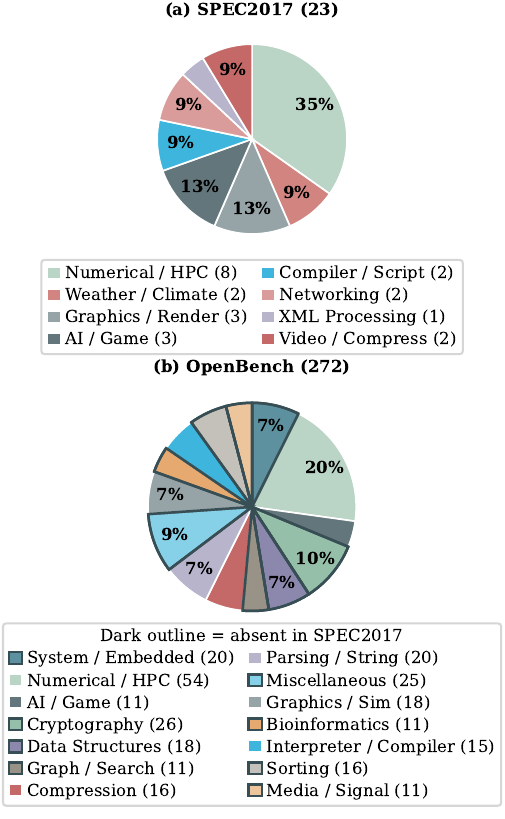}
    \caption{Application-domain distribution of \textit{openbench-universe} and SPEC CPU2017. The pie chart illustrates that \textit{openbench-universe} encompasses a significantly broader range of modern application domains than SPEC CPU2017, reflecting contemporary software ecosystems.}
      \Description{Pie charts comparing application domain distributions of openbench-universe and SPEC CPU2017, showing broader domain diversity in openbench-universe.}
    \label{fig:domain_dist}
\end{figure}

\setcounter{dbltopnumber}{1}
\begin{figure*}[t]
    \centering
    \includegraphics[width=\textwidth]{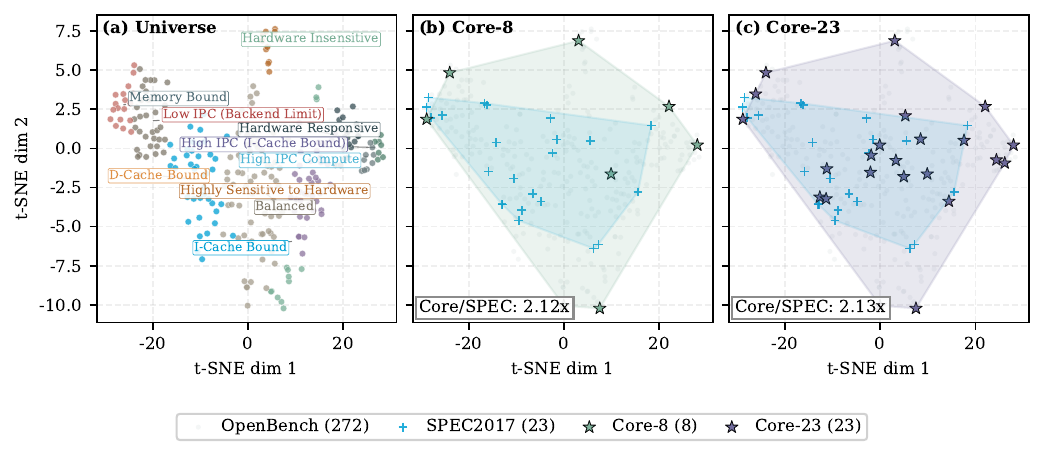}
    \caption{Unified t-SNE projection of the \textit{openbench} workload space across C1--C30, illustrating the relationship between the full universe and its distilled subsets. All three panels share the same embedding.
    \textbf{(a)}~The full 272-workload \textit{openbench-universe} grouped into 10 behavioral clusters via K-Means. Each cluster is annotated with an intuitive microarchitectural archetype, ranging from \emph{Memory Bound} to \emph{High IPC (I-Cache Bound)}, reflecting its dominant hardware bottleneck.
    \textbf{(b)}~\textit{openbench-core} Core-8 (green stars) versus SPEC2017 (cyan crosses). The Core-8 convex hull ($2.12\times$ SPEC hull area) substantially exceeds the SPEC2017 footprint with only eight benchmarks.
    \textbf{(c)}~\textit{openbench-core} Core-23 (purple stars) versus SPEC2017 (cyan crosses). Core-23 is a \emph{nested superset} of Core-8 (Core-8 $\subset$ Core-23), so its hull provably contains the Core-8 hull; it reaches $2.13\times$ the SPEC hull area while further improving fidelity to the full universe.}
    \Description{Three-panel t-SNE scatter plot: (a) openbench-universe with 10 K-Means clusters annotated by microarchitectural trait; (b) Core-8 vs SPEC2017 with convex hulls; (c) Core-23 vs SPEC2017 with convex hulls.}
    \label{fig:clustering_pca}
\end{figure*}

As shown in Figure~\ref{fig:domain_dist}, SPEC2017 concentrates on a narrow set of traditional domains, whereas the 272 workloads in \textit{openbench-universe} cover a drastically wider variety of modern domains including machine learning, graph processing, and cryptography. Moreover, whereas all 23 SPEC CPU2017 benchmarks are written in legacy languages such as C, C++, and Fortran, \textit{openbench-universe} incorporates applications written in Rust, Go, and other modern systems languages, capturing the distinct execution characteristics of contemporary software stacks.

Beyond application-level breadth, Figure~\ref{fig:clustering_pca} reveals the \emph{microarchitectural} diversity of \textit{openbench-universe}. We represent each workload as a 30-dimensional IPC feature vector spanning C1--C30, project it to two dimensions via t-SNE, and apply K-Means clustering with $k{=}10$. Rather than labeling clusters by application domain, we assign each an intuitive \emph{behavioral archetype} from its aggregate microarchitectural profile, including mean IPC, cache miss rates, and configuration sensitivity.

The projection confirms that \textit{openbench} workloads do not collapse into a monolithic group but disperse into ten well-separated clusters spanning the full spectrum of hardware bottlenecks. At one extreme lies the \emph{Memory Bound} cluster, whose members achieve the lowest IPC due to severe data-cache pressure. At the other extreme, the \emph{High IPC (I-Cache Bound)} cluster sustains the highest throughput despite elevated instruction-fetch miss rates. Between these poles, \emph{D-Cache Bound} and \emph{Low IPC (Backend Limit)} workloads expose distinct memory-hierarchy pathologies, while \emph{Hardware Insensitive} benchmarks remain indifferent to configuration changes. The \emph{Highly Sensitive to Hardware} and \emph{Hardware Responsive} clusters exhibit pronounced IPC variation across the 30 configurations, making them valuable for stress-testing pipeline and cache trade-offs. The \emph{I-Cache Bound} cluster shares elevated instruction-cache miss rates with its High IPC counterpart yet operates at moderate throughput, isolating pure frontend pressure, and the \emph{High IPC Compute} cluster represents workloads that are primarily compute-limited. The largest cluster, \emph{Balanced}, anchors the center of the embedding with moderate, well-distributed resource demands.

Every cluster carries a unique label because the ten archetypes derive from multi-dimensional profiles rather than any single scalar metric: two clusters sharing a similar mean IPC still diverge along orthogonal axes such as cache-miss distribution and configuration sensitivity. By exposing this broad behavioral landscape, \textit{openbench} gives architects the bounds to stress-test robust processor designs against the diversity of modern software.

\begin{figure*}[t]
    \centering
    \includegraphics[width=\textwidth]{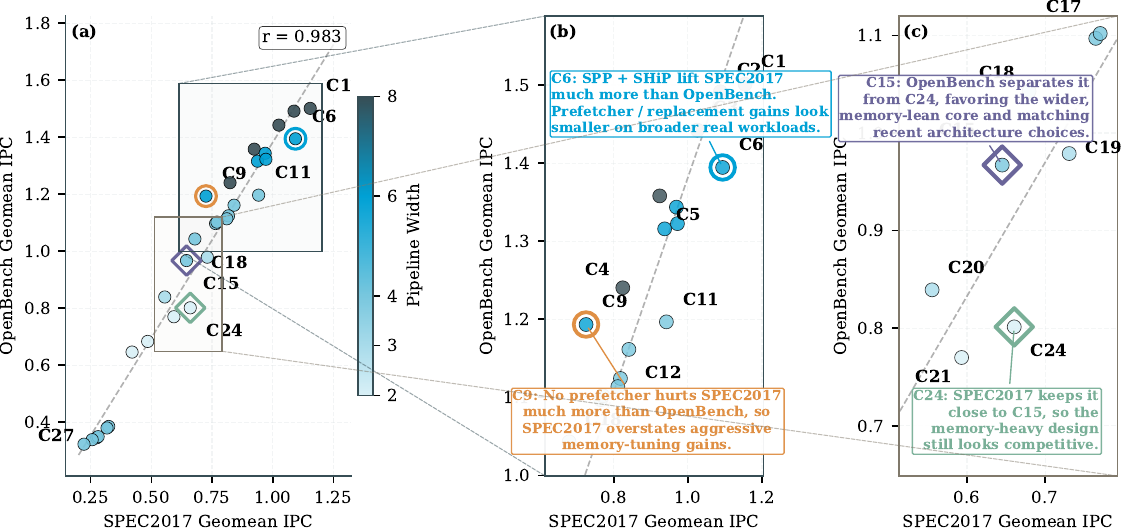}
    \caption{Geometric-mean IPC of \textit{openbench} versus SPEC2017 across 30 microarchitectural configurations (C1--C30), colored by pipeline width. \textbf{(a)}~Full scatter plot showing the strong linear correlation ($r = 0.983$), while notable outliers reveal configuration-specific divergences invisible to either suite alone. \textbf{(b)}~Zoomed view detailing the two most off-trend configurations. The different responses of \textbf{C9} and \textbf{C6} show that aggressive prefetcher and cache-replacement choices appear more effective on SPEC2017 than on \textit{openbench}, suggesting that these memory optimizations are partly overfit to SPEC2017 and deliver smaller gains on broader modern workloads. \textbf{(c)}~Zoomed view of \textbf{C24} (2-wide, memory-heavy: deep ROB, 1\,MB L2, SPP prefetcher) versus \textbf{C15} (4-wide, memory-lean: shallow ROB, 256\,kB L2, no prefetcher). SPEC2017 treats the two designs as nearly tied, whereas \textit{openbench} separates them and favors C15, revealing a design distinction that SPEC2017 alone obscures and making \textit{openbench} more informative for architecture evaluation and selection.}
      \Description{Three-panel scatter plot: (a) full view of all 30 configurations with regression line; (b) zoomed view of configurations C9 and C6 showing that prefetcher and cache-replacement gains appear larger on SPEC2017 than on openbench; (c) zoomed view of C24 and C15 showing that SPEC2017 treats them as nearly tied while openbench clearly favors C15.}
    \label{fig:scatter_plot_universe}
\end{figure*}

To further analyze the difference between \textit{openbench-universe} and SPEC2017, Figure~\ref{fig:scatter_plot_universe} plots the \textit{openbench} geometric-mean IPC against the SPEC2017 geometric-mean IPC across all 30 configurations. The strong Pearson correlation of $r = 0.983$ confirms suite-level consistency, yet notable outliers expose configuration-specific divergences, and Figure~\ref{fig:scatter_plot_universe}(b) zooms into the two most off-trend configurations. Configuration C9, a 6-wide design without a prefetcher, achieves a disproportionately high \textit{openbench} IPC relative to its SPEC2017 score because the absent prefetcher penalizes SPEC2017's cache-sensitive workloads more severely while \textit{openbench}'s diverse applications are comparatively resilient. Conversely, C6, which pairs a 64\,kB L1 with an SPP prefetcher and SHiP replacement, falls below the regression line: its advanced cache-hierarchy optimizations substantially boost SPEC2017's regular access patterns but provide much smaller gains on \textit{openbench}'s heterogeneous mix. Together, these two points indicate that modern prefetcher and cache-replacement designs can look stronger on SPEC2017 than on broader real workloads, suggesting a degree of overfitting to SPEC2017 that \textit{openbench} makes visible.

Beyond individual outliers, Figure~\ref{fig:scatter_plot_universe}(c) illustrates a systemic design trade-off by zooming into C24 and C15. C24 is a 2-wide core with a heavily provisioned memory subsystem: a deep 128-entry reorder buffer, a large 1\,MB L2, and an aggressive Signature Path Prefetcher with DRRIP replacement. C15 is a wider 4-wide core with a lean memory subsystem: a shallow 96-entry reorder buffer, a compact 256\,kB L2, and no hardware prefetcher. Under SPEC2017, C24 and C15 achieve parity. The \textit{openbench-universe} evaluation instead reveals an inversion where the lean-memory C15 significantly outperforms C24: on modern applications with massive code footprints, intricate branch behavior, and irregular memory access, C15's wider pipeline sustains higher throughput while C24's deep memory resources stay underutilized amid frequent stalls. C15 and C24 are deliberately unbalanced corner designs; the point is not that C15 is universally better, but that benchmark choice alone can flip the apparent ranking of two candidates. Relying solely on SPEC2017 can thus obscure meaningful architectural differences during design-space exploration, and complementing it with \textit{openbench} restores visibility into these distinctions.

\begin{figure}
    \centering
    \includegraphics[width=3.5in]{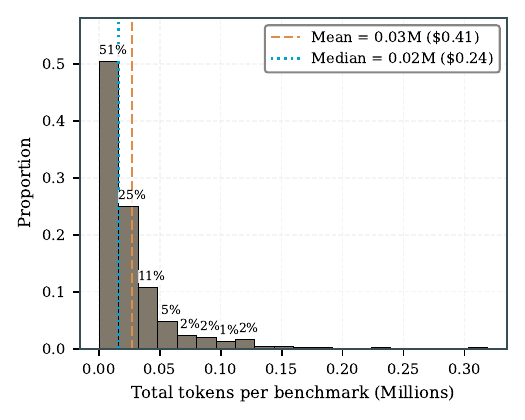}
    \caption{Token-usage distribution of \textit{openbench-universe} under MiniMax-M2.5 in millions of tokens. Most benchmarks fall below 0.05M tokens. The mean token usage is 0.03M and the median token usage is 0.021M, showing the extremely low cost and scalability of \textit{BenchmarkAnything}.}
    \Description{Histogram showing the token-usage distribution of openbench-universe benchmarks, with most values below 0.05M tokens.}
    \label{fig:token_usage_hist}
\end{figure}

To evaluate the overhead of constructing \textit{openbench-universe}, we analyze token usage. Figure~\ref{fig:token_usage_hist} shows the distribution: most benchmarks fall below 0.05M tokens, with a mean of 0.03M and a median of 0.021M, demonstrating the extremely low cost and scalability of \textit{BenchmarkAnything}.
\subsection{Experiment-based Selection from \textit{openbench-universe} to \textit{openbench-core}}
While \textit{openbench-universe} provides unprecedented scale, practical architectural simulation demands compact, non-redundant suites. We therefore apply a principled statistical distillation to reduce the 272 raw workloads into a concise representative suite, denoted \emph{\textit{openbench-core}}.

\begin{figure}[tbp]
    \centering
    \includegraphics[width=3.5in]{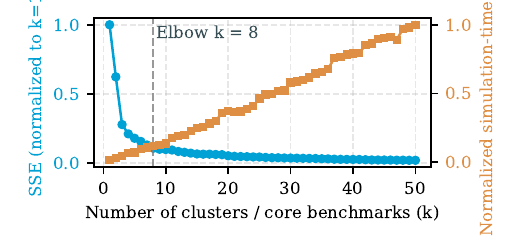}
\caption{Determining the optimal number of core benchmarks ($k$). The left axis denotes the SSE normalized to $k=1$. The right axis displays a normalized simulation-cost proxy computed by summing the selected medoids' \texttt{simSeconds} measured under baseline configuration C30 and then normalizing by the largest evaluated subset size. K-medoids clustering in the PCA space reveals an elbow at $k=8$, offering a strong trade-off between representative accuracy and simulation efficiency.}
  \Description{Elbow plot showing normalized SSE and a normalized simulation-cost proxy versus the number of clusters k, with a clear inflection at k equals 8.}
  \label{fig:elbow_detection}
\end{figure}

We represent each benchmark as a 30-dimensional IPC feature vector spanning C1--C30 and apply PCA to project this space onto its principal components. Figure~\ref{fig:elbow_detection} shows the elbow analysis that determines the optimal subset size: k-medoids clustering in the PCA feature space reveals a clear inflection at $k=8$, providing a strong trade-off between representational fidelity and simulation cost.

With $k$ determined, we run k-medoids on the Pearson-distance matrix in PCA space to select one representative medoid per cluster. We adopt Pearson rather than Euclidean distance because it measures the similarity of IPC \emph{profiles} across configurations, capturing how a benchmark responds to microarchitectural changes irrespective of absolute throughput. Two workloads with identical scaling trends but different absolute IPC stress the same hardware structures in the same relative way and should share a cluster, whereas Euclidean distance would conflate magnitude differences with profile-shape differences. Each medoid is an actual benchmark whose per-configuration IPC profile lies closest to its cluster centroid in the reduced feature space. To prevent the selection of obscure micro-benchmarks, we augment the distance-based objective with a soft tie-breaking prior that favors widely adopted projects.

Table~\ref{tab:openbench-core8} lists the resulting Core-8 subset in \textit{openbench-core}. Each row corresponds to one medoid retained from a distinct behavioral cluster, and the justification column summarizes the dominant property that made that benchmark the representative of its cluster. The eight benchmarks also span eight distinct application domains, from memory-latency stress tests and hash-intensive data structures to bioinformatics alignment and physics simulation, ensuring broad algorithmic diversity in addition to microarchitectural coverage. To prove that it is a representative subset of \textit{openbench-universe}, we validate Core-8 along two complementary axes:

\begin{table}[t]
  \centering
  \caption{\textit{openbench-core} Core-8 benchmark set selected by PCA+k-medoids. Each medoid represents one behavioral cluster and is chosen to maximize coverage.}
  \label{tab:openbench-core8}
  \footnotesize
  \setlength{\tabcolsep}{3pt}
  \begin{tabularx}{\columnwidth}{@{}c>{\raggedright\arraybackslash}p{1.35cm}>{\raggedright\arraybackslash}p{1.55cm}X@{}}
    \toprule
    ID & Benchmark & Domain & Justification \\
    \midrule
    1 & mlc-like & System / Embedded & Memory latency stress test \\
    2 & hashbrown & Data Structures & Rust std HashMap, hash-intensive \\
    3 & stb-bench & Media / Signal & Single-header image codec \\
    4 & mbedtls & Cryptography & Widely-deployed TLS library \\
    5 & box2d-v3 & Graphics / Sim & Industry-standard physics engine \\
    6 & minimap2 & Bioinformatics & Reference genome aligner \\
    7 & Fairy-Stockfish & AI / Game & Chess variant engine, tree search \\
    8 & HPCCG & Numerical / HPC & DOE conjugate-gradient proxy app \\
    \bottomrule
  \end{tabularx}
\end{table}

\begin{figure}[t]
    \centering
    \includegraphics[width=3.5in]{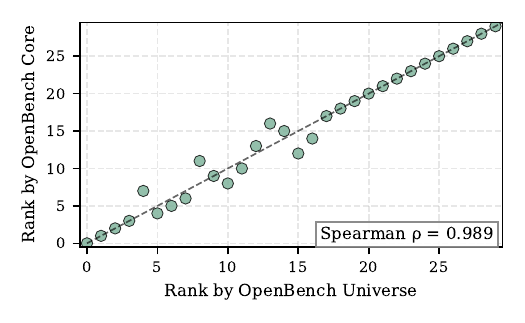}
\caption{Correlation between configuration rankings. The x-axis represents the rank by the entire \textit{openbench-universe}. The y-axis represents the rank by the \textit{openbench-core} subset. Both rankings evaluate the geometric mean of IPC with the best configuration ranked as 0. The evaluation yields a Spearman $\rho=0.989$. This correlation confirms that hardware design decisions made using the 8-benchmark core subset closely mirror those made using the full benchmark suite.}
  \Description{Scatter plot showing the strong correlation between configuration rankings from the full openbench-universe and from the openbench-core subset, with Spearman rho of 0.989.}
  \label{fig:rank_correlation}
\end{figure}

\textit{Ranking fidelity.}  Figure~\ref{fig:rank_correlation} compares configuration rankings from Core-8 against the full universe, both ordered by geometric-mean IPC. The Spearman coefficient $\rho = 0.989$ confirms that design decisions based on the eight-benchmark subset closely mirror those from all 272 workloads.

\begin{figure}[!t]
    \centering
    \includegraphics[width=3.5in]{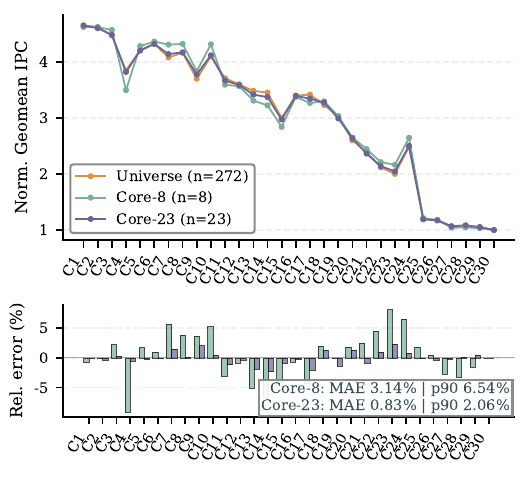}
    \caption{Performance fidelity of the core subsets compared to the \textit{openbench-universe} across 30 configurations (C1--C30). The top panel overlays the normalized geometric-mean IPC trends for the full universe, Core-8, and Core-23. The bottom panel reports each subset's relative error versus the full universe. Core-8 achieves mean absolute / $p90$ / max errors of $3.14\%$ / $6.54\%$ / $9.10\%$, while Core-23 tightens these bounds to $0.83\%$ / $2.06\%$ / $2.31\%$.}
    \Description{Single-column figure showing normalized IPC trends for the openbench-universe, Core-8, and Core-23 across configurations, plus a grouped relative-error bar chart for Core-8 and Core-23.}
    \label{fig:core_vs_openbench_total_score}
\end{figure}

\textit{Performance fidelity.}  Figure~\ref{fig:core_vs_openbench_total_score} overlays normalized geometric-mean IPC curves for the core subsets and the full universe. Core-8 achieves a mean absolute error of $3.14\%$, a $p_{90}$ error of $6.54\%$, and a maximum error of $9.10\%$. For a direct, equal-cardinality comparison with the 23 SPEC CPU2017 benchmarks, we also construct Core-23 as a \emph{nested superset} of Core-8: we retain all eight Core-8 medoids and greedily add fifteen further benchmarks, each chosen to most reduce the residual error against the full universe. Because Core-8 $\subset$ Core-23, every set-monotone coverage measure (including the convex hull of Sec.~\ref{Characterization}) is guaranteed not to shrink as the subset grows, removing the counterintuitive possibility that a smaller subset covers more of the behavioral space than a larger one. Core-23 tightens these bounds to $0.83\%$, $2.06\%$, and $2.31\%$, respectively. The lower panel further shows that the Core-23 residual bars are smaller than Core-8's for most configurations and lower in aggregate, confirming that the larger subset preserves configuration-specific behavior more faithfully rather than only matching the average trend.

\begin{figure*}[t]
  \centering
  \includegraphics[width=1\textwidth]{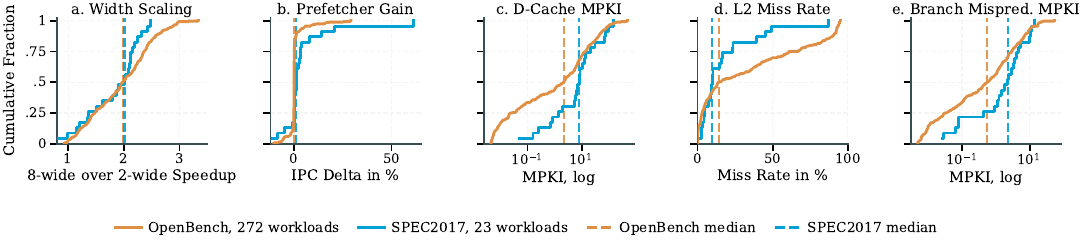}
  \caption{Microarchitectural characteristics of \textit{openbench} and SPEC CPU2017. Every panel shows one cumulative distribution per suite, and dashed vertical lines mark per-suite medians. Panel a reports the IPC speedup from the 2-wide C24 to the 8-wide C1: the medians agree near 2.0, yet 47 of 272 \textit{openbench} workloads exceed the SPEC2017 maximum of 2.5. Panel b reports the IPC delta between SPP with DRRIP in C1 and a stride prefetcher with SHiP in C2: returns are consistently positive on SPEC2017 but near zero for the median \textit{openbench} workload, so prefetching gains measured on SPEC2017 overstate what broader workloads receive. Panels c to e report hardware event rates on the mainstream reference core C12, recomputed from the same gem5 runs: \textit{openbench} extends the data-side envelope, with a D-cache MPKI tail reaching 470 versus 141 for SPEC2017, an L2 miss rate tail reaching 95\,\% with 39 workloads above the SPEC2017 maximum, and a branch misprediction MPKI tail reaching 53 versus 14. The two suites are complementary rather than interchangeable.}
  \Description{Five-panel figure of cumulative distributions comparing openbench and SPEC2017: pipeline width speedup, prefetcher gain, D-cache MPKI, L2 miss rate, and branch misprediction MPKI. Dashed lines mark per-suite medians; openbench extends the data-side envelope while the suite medians remain comparable.}
  \label{fig:arch_overview}
\end{figure*}

Finally, we compare \textit{openbench-core} against SPEC2017 in the shared t-SNE embedding. As shown in Figure~\ref{fig:clustering_pca}(b) and~(c), Core-8 achieves $2.12\times$ the SPEC2017 convex-hull area; because Core-23 is a nested superset of Core-8, its hull necessarily contains the Core-8 hull and therefore covers at least as much of the behavioral space, reaching $2.13\times$. Even small distilled subsets thus preserve much of the broad microarchitectural diversity of the full universe while dramatically reducing simulation cost.

\section{Discussion}
Our analysis reveals a recurring pattern across a broad set of microarchitectural structures.
As shown in Figure~\ref{fig:arch_overview}, optimizations that appear highly effective on SPEC2017 often generalize less reliably to the more diverse \textit{openbench-universe}.
Every panel uses the same visual grammar: a cumulative distribution over the workloads of each suite with dashed per-suite median lines.
Panel a shows that the two suites agree on the median width speedup near 2.0, but only \textit{openbench} exposes a strongly width-sensitive tail, with 47 of 272 workloads above the SPEC2017 maximum.
Panel b shows that prefetcher and replacement gains are consistently positive on SPEC2017 while the \textit{openbench} median sits near zero: returns measured on the legacy suite generalize to fewer modern workloads.
Panels c to e move beyond IPC and report hardware event rates: D-cache misses per kilo instruction, L2 miss rate as a proxy for memory pressure and working set size, and branch misprediction rate.

These event distributions expose exactly the information an IPC-only view can hide: \textit{openbench} extends the envelope on D-cache MPKI, L2 miss rate, and branch misprediction MPKI, with 39 workloads exceeding the SPEC2017 maximum L2 miss rate and 12 exceeding its maximum branch misprediction MPKI.
\textit{openbench} thus broadens memory-hierarchy and control-flow coverage rather than uniformly shifting every metric, reflecting its mix of compact compute kernels and a heavy memory-bound tail, so the two suites are complementary rather than interchangeable.

These discrepancies suggest that contemporary CPU microarchitectures may be partially over-tuned to legacy benchmark behavior.
For decades, the community has used SPEC suites as a primary metric for academic and industrial success, so branch predictors, cache replacement policies, and data prefetchers may have been disproportionately optimized for the regular access patterns that dominate these legacy workloads.
This creates a feedback loop in which hardware mechanisms are designed to maximize SPEC performance and are then validated on the same suite.
The diminished returns observed on \textit{openbench} suggest that some highly regarded optimizations generalize less reliably to the heterogeneous execution behaviors of modern software ecosystems.

We therefore raise an important question for the architecture community: how much of our established microarchitectural intuition reflects general performance truth, and how much reflects historical benchmarking bias?
Future design studies should complement legacy suites with broader and continuously refreshed workload collections, a direction that agent-driven frameworks such as \textit{BenchmarkAnything} make practical through large-scale workload acquisition, standardization, and distillation.

Our current evaluation also has clear bounds.
A primary limitation is the use of gem5 in syscall-emulation mode, which abstracts away kernel execution, page faults, context switches, and complex input-output behavior.
Workloads that depend heavily on operating-system activity or network stacks may therefore not expose their full bottlenecks in the current setup.

These bounds are properties of the \emph{current instance} of OpenBench, not of the \textit{BenchmarkAnything} methodology. Today's instance targets single-threaded, user-level, statically linked, AVX-free x86 binaries because that is the envelope gem5 SE mode accepts; this scope prunes some application classes, such as multithreaded, network-intensive, or OS-intensive software and AVX-512-heavy kernels, introducing a degree of selection bias. Every such constraint, however, lives in the build- and validation-rule set given to the agents, not in the agent flow itself. Retargeting the same loop to gem5 full-system mode, ARM or RISC-V, AVX-enabled rebuilds, or multithreaded workloads requires changing those declarative constraints rather than redesigning the framework; the methodology is portable even though the present artifact is deliberately scoped for reproducibility.

Future work will extend \textit{BenchmarkAnything} in three directions: full-system simulation so that agent-generated workloads capture richer operating-system behavior, agent workflows that synthesize and compile multi-threaded applications, and heterogeneous targets including GPUs and domain-specific accelerators. Together, these extensions will broaden \textit{openbench-universe} into a more comprehensive evaluation platform for future system architectures.

\section{Conclusion}
Evaluating modern processors requires workloads that accurately reflect contemporary software. To overcome the limitations of manual benchmark curation, we presented \textit{BenchmarkAnything}, an automated agent-driven methodology. This framework synthesizes \textit{openbench-universe}, a suite of 272 applications collected from open-source platforms, and distills it into \textit{openbench-core}, a highly efficient representative subset. Our evaluation demonstrates that Core-8 in \textit{openbench-core} achieves up to $2.12\times$ the behavioral-space coverage of SPEC CPU2017 while preserving strong ranking and performance fidelity to the full universe. By reducing human selection bias and exposing critical execution bottlenecks missed by legacy suites, \textit{openbench} provides a compelling target for designing more robust future architectures.


\bibliographystyle{IEEEtran}
\bibliography{refs}

\end{document}